\documentclass{nime-alternate}

\begin{document}
%
\conferenceinfo{NIME'13,}{May 27 -- 30, 2013, KAIST, Daejeon, Korea.}

\title{Enactive Mandala: Audio-visualizing Brain Waves}

\numberofauthors{2} 
\author{
\alignauthor
Tomohiro Tokunaga\\
      \affaddr{Ritsumeikan University}\\
       \affaddr{Kyoto, Japan}\\
       \email{byorori.t@gmail.com}
\alignauthor
Michael J. Lyons\\
       \affaddr{Ritsumeikan University}\\
       \affaddr{Kyoto, Japan}\\
       \email{michael.lyons@gmail.com}
}

\maketitle
\begin{abstract}
We are exploring the design and implementation of artificial expressions,
kinetic audio-visual representations of real-time physiological data which reflect
emotional and cognitive state.  In this work we demonstrate a prototype, the \emph{Enactive Mandala}, which maps real-time EEG signals to modulate ambient music and animated visual music. Transparent real-time audio-visual feedback of brainwave qualities supports intuitive insight into the connection between thoughts and physiological states. 
\end{abstract}


\section{Introduction}
Some art forms share strategies with meditative practices for experiencing, from a fresh perspective, experience itself. Interactive media art, with the potential to interface to the human mind and body at previously impossible levels of intimacy, presents artists unparalleled opportunities for creating new perspectives on the subjective experience of the embodied mind. In this work our aim is to design an interactive audio-visual (a/v) system to augment perception of the relationship between mental and physiological state and the influence one's mental state may have on the external world. 

Sonifying electrical signals from the activity of human brains has a long and rich tradition in electronic and computer music. In recent years advances in dry electrode technology has led to the development of EEG (electroencephalography) devices that enable the measurement of brainwaves without the need to apply conductive gel to the scalp \cite{Neurosky:2009}. Together with a significant drop in the cost of these systems, this has largely reduced technical and financial barriers to getting involved in research on brainwaves. It is important to recognize that EEG signals, the summed result of massively complex electrical activity in the brain, are not completely understood in terms of how they relate to the details of the activity of neural circuits, let alone mental or affective states. Decades of research, however, have shown meaningful correlations with behaviour and self-reported mental state \cite{austin06}.

Here we  explore a constructive approach to understanding brainwave data in the context of musical/artistic expression. Real-time EEG data is mapped to visual and auditory outputs. Subjective states can become intuitively associated with the a/v displays, because there is little lag between measurement of the EEG and its a/v representation. The EEG a/v output then serves as a mirror and physiological correlate of mental activity upon which to ground reflection on one's subjective experience of the present moment. Insight is gained \emph{enactively} \cite{varela1991} as a user implicitly observes an intuitive connection between internal thoughts and the state of an interactive a/v mandala.

\section{Method}
Our approach is directly related to  two prior works. Lyons et al. \cite{Lyons:2004} proposed the general concept of  \emph{artificial expressions}  (AE), novel machine-mediated forms of non-verbal communication based on real-time physiological data. A key component of an AE is an effective mapping of physiological data to visible or audible gestalt to create ambient real-time non-verbal cues which can be understood implicitly. Deploying the AE long-term in meaningful communicative situations leads inter-actors to construct meanings for the AE as they experience it in the context of changing circumstances. Barrington et al. \cite{Barrington:2006} extended this approach to the auditory domain, applying beat-synced audio effects to a playlist of music tracks, thus adding a communicative layer of AE. These  ``affective effects'' were used to prototype a calm auditory notification system \cite{Barrington:2006}. The current work combines audio and visual approaches. 
\begin{figure}
	\centering
		\includegraphics[width=1\columnwidth]{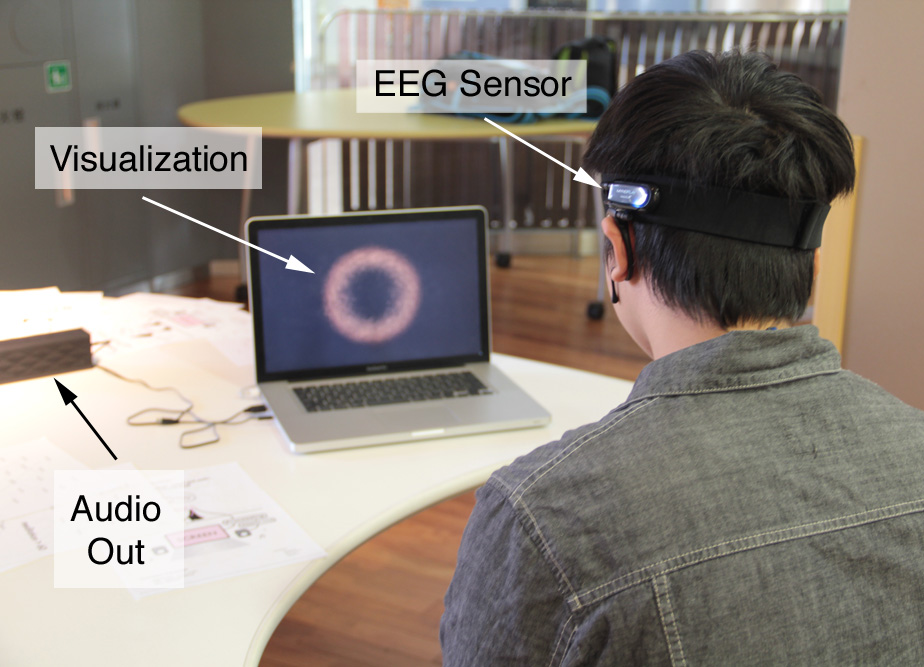}
	\caption{Enactive Mandala Demonstration}
	\label{fig:fig1}
\end{figure}

\begin{figure*}
	\centering
		\includegraphics[width=1\textwidth]{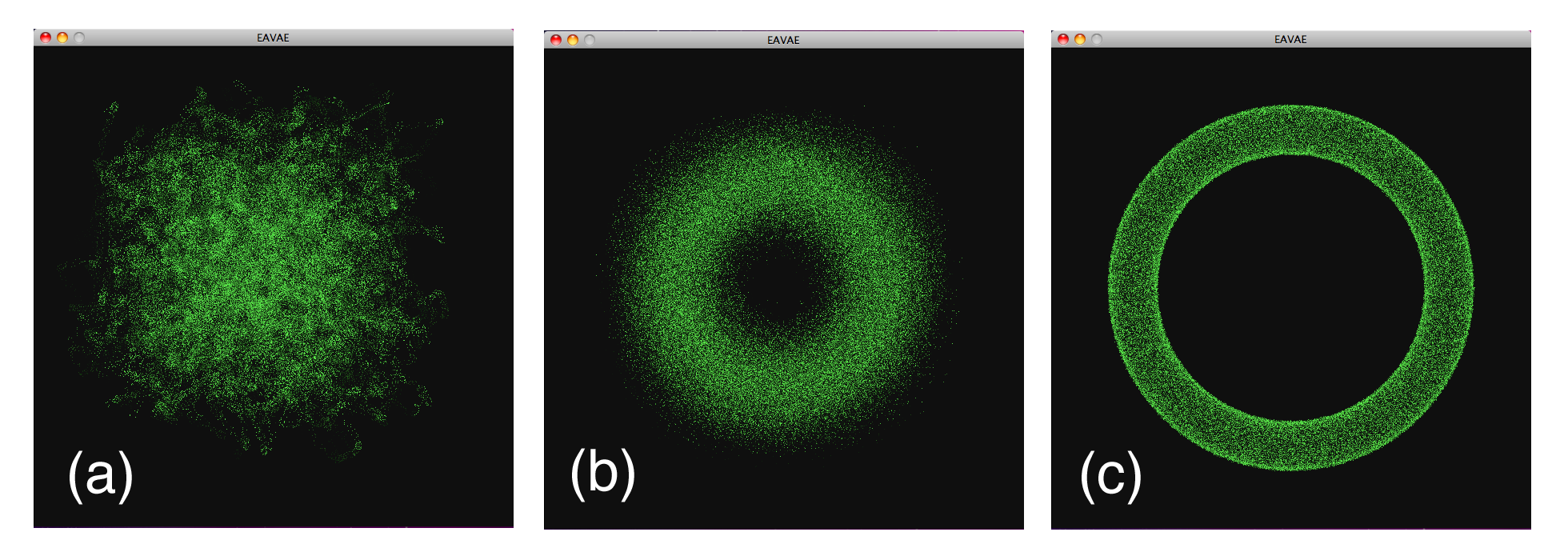}
	\caption{Visualization for (a) low values of M (b) intermediate M, and (c) high M.}
	\label{fig:fig2}
\end{figure*}
A NeuroSky MindSet is used to obtain several components of the EEG signal in real time with the  \emph{np\_mindset} Max/MSP external.\footnote{http://www.github.com/qdot/np\_mindset}
The MindSet outputs composite brainwave measures described as Meditation (M) and Attention (A) at 1 Hz. NeuroSky has published correlations with EEG components as evidence that these reflect mental relaxation and alertness \cite{Neurosky:2009}.

For the current work we primarily make use  of the M signal. Taking inspiration from some traditional mandala designs\footnote{Mandala in Sanskrit literally means `circle'.} and the visual music of the Whitney brothers \cite{Whitney:1981}, the EEG signal, M, is visualized by a system of $L$ particles moving epicyclically (Figure \ref{fig:fig2}). Particle $q$ moves in the complex plane $z = (x, iy)$ according to:
\begin{equation*}
z_{q}(t-t_{q}) = M(Re^{i\omega_{q} t} - re^{i\Omega_{q} t}) + (1-M)(N_{qx}+iN_{qy})\\
\end{equation*}
Radii $R, r$, angular speeds $\omega, \Omega$, and phase factor $t_{q}$ are real-valued constants. The $N_{qx,qy}$ are $2L$ uncorrelated Perlin noise processes. As M decreases, noise disrupts the coherence of the epicycles, dissolving the circular ring. This shape somewhat resembles the \emph{Ens\={o}} figure of Zen Buddhism, but it is not intended as a replica of any specific traditional mandala. For the audio component, M is sent via the OSC protocol to Ableton Live or Max/MSP to cross-fade between tracks $T_{1}$ and $T_{2}$:
\begin{align*}
S(t) = MT_{1}(t) + (1-M)T_{2}(t)
\end{align*}
For example, $T_{1}$ may be a harmonious ambient loop of layered guitar, and $T_{2}$ a jumble of staccato, noisy samples to reflect quiet and perturbed mental states. Cross-fading applies whether the audio is mono, stereo, or multi-channel.
A disadvantage of the system is that it requires custom audio loops.To add flexibility, we use the strategy \cite{Barrington:2006} of applying audio effects to a track playlist.
To give one example,  the constant-pitch playback rate $R(t)$ of an audio file is modulated via granular synthesis:
\begin{align*}
R(t) = MR_{0}, \;\; \Delta \tau = \alpha (1-M)n(t)
\end{align*}
For M = 1, the audio track is played at the normal rate $R_{0}$, but as M drops, the playback rate decreases. Sample position is changed stochastically by $\Delta \tau$, with $\alpha$ a constant to set the magnitude of uniform noise process $n(t)$. For M=0, one hears a disordered jumble of sound particles, a sonic analog of panel (a) in Figure \ref{fig:fig2}.

The first functioning version (Figure \ref{fig:fig1}) of the system was demonstrated within our department at the ``Eizo Junction'' event in December 2011. An immersive installation with large screen, 4.1 audio (4 surround speakers a sub-woofer) in a quiet isolated space was demonstrated in December 2012. In Autumn 2012 we conducted carefully controlled experiments with 20 subjects in which the system was compared with one where the a/v mapping was reversed. A detailed analysis of the findings is beyond the scope of this demo abstract and will be reported elsewhere. In brief, objective measures showed significantly superior performance with the forward design.  Subjective measures showed users perceive a causal link between their mental state and the a/v display for both forward and reverse conditions.
\section{Outlook}
We have designed and implemented a system to explore the construction of new modes of expressive communication aimed at inducing a novel artistic and musical experience reflecting mental state. The system takes EEG features as input and uses these to drive real-time visual and auditory analogs of the captured data. Our method is constructive: by linking physiology with an dynamic a/v display, and embedding the human-machine system in the social contexts that arise in real-time play, we hope to seed new, and as yet unknown forms, of non-verbal communication, or ``artificial expressions''. 
\bibliographystyle{abbrv}
\bibliography{nime13_ae} 
\balancecolumns 

\end{document}